# THE EAP-AIAS: ADAPTING THE AI ASSESSMENT SCALE FOR ENGLISH FOR ACADEMIC PURPOSES

Jasper Roe [1*], Mike Perkins [2], Yulia Tregubova [2]

[1] James Cook University Singapore, Singapore
[2] British University Vietnam, Vietnam.
[*] Corresponding Author: jasper.roe@jcu.edu.au

August 2024

## Abstract

The rapid advancement of Generative Artificial Intelligence (GenAI) presents both opportunities and challenges for English for Academic Purposes (EAP) instruction. This paper proposes an adaptation of the AI Assessment Scale (AIAS) specifically tailored for EAP contexts, termed the EAP-AIAS.

This framework aims to provide a structured approach for integrating GenAI tools into EAP assessment practices while maintaining academic integrity and supporting language development. The EAP-AIAS consists of five levels, ranging from "No AI" to "Full AI", each delineating appropriate GenAI usage in EAP tasks. We discuss the rationale behind this adaptation, considering the unique needs of language learners and the dual focus of EAP on language proficiency and academic acculturation.

This paper explores potential applications of the EAP-AIAS across various EAP assessment types, including writing tasks, presentations, and research projects. By offering a flexible framework, the EAP-AIAS seeks to empower EAP practitioners seeking to deal with the complexities of GenAI integration in education and prepare students for an AI-enhanced academic and professional future. This adaptation represents a step towards addressing the pressing need for ethical and pedagogically sound AI integration in language education.

# Introduction

Advancements in the field of Artificial Intelligence (AI) and Generative AI (GenAI) have sparked intense debate across educational disciplines, with the release of sophisticated AI models such as OpenAI's ChatGPT and Anthropic's Claude representing a shift in the capabilities of language technologies, which shows no sign of slowing down. At the time of writing, such tools now encompass multimodal elements, voice generation, and the ability to produce instant and increasingly sophisticated responses that can closely mimic human-like communication. These technologies are poised to become smaller, more cost-efficient, localised, and easily downloadable, meaning that they may soon become an available tool to everyone who owns a smartphone or other communication device, regardless of whether the device has access to the internet.

In educational research, on one end of the spectrum are those who believe that AI and GenAI represent vague, problematic terms for a set of technologies which are fundamentally about commercial gain, rather than improving the educational process, and bring more harm to the practice of English language teaching than good (Caplan, 2024). Indeed, it has been asked whether these technologies are just 'bullshit spewers' (Rudolph et al., 2023) or merely produce 'botshit' (coherent sounding, counterfactual content) (Hannigan et al., 2024). On the other hand, surveys among students and faculty generally reveal positive attitudes towards these technologies, with some caution and scepticism (Chan & Hu, 2023; Roe et al., 2024). Among the general population, surveys demonstrate that many feel that GenAI represents a 'world-shifting, permanent, step-change in technology and media', evidenced by the app ChatGPT's ability to garner over 100 million users in as little as two months, shattering previous records (Archer, 2024).

Despite this attention on both ends of the spectrum, GenAI's role in English for Academic Purposes (EAP) remains a relatively unexplored area (Kostka & Toncelli, 2023); however, EAP faces unique and specific challenges in this area due to its dual focus on language proficiency and academic acculturation. EAP is at risk as there is the potential for students to use GenAI tools to disguise authorship, complete tasks without genuine engagement, or rely on them for automated machine translation of texts and even classroom discussion, thereby potentially circumventing the learning processes that EAP aims to facilitate. In the context of EAP assessment specifically, these concerns are acute because of the prevalence of take-home assessments in EAP practice, coupled with the broad focus of EAP assessments assessing not only content but also vocabulary diversity and choice, grammatical range and accuracy, coherence, structure, and other aspects of language proficiency and written expression.

Regarding take-home assessments, although proctored examinations are common, EAP courses often encompass take-home written assignments, portfolios, and presentations as core components of assessments, as they are authentic and mimic tasks required in later academic studies. These formats, while valuable for developing academic skills, are particularly vulnerable to GenAI-assisted completion or outright AI generation, as tools such as ChatGPT can be used to both fix errors in writing and compose essays (Godwin-Jones, 2022; Perkins, 2023). As mentioned above, EAP assessments also evaluate not only the content presented but also the linguistic elements, including grammatical and lexical accuracy, structure, and genre-specific stylistic expressions. Therefore, the ability of GenAI tools to produce grammatically flawless, well-structured, and sophisticated writing poses a significant challenge to this assessment model. Additionally, EAP assessments often target a range of skills that go beyond a student's linguistic ability, such as critical thinking, academic, and research skills. In a pre-GenAI era, a response essay, for example, would require a student to take several steps involving a certain degree of critical engagement before the actual writing – selecting credible sources, reading and critically engaging with their content, identifying similarities and differences in the views or approaches, synthesising, paraphrasing, and summarising, and then creating their own work. With the assistance of GenAI, however, it is possible to proceed from step one directly to the final step, which potentially not only deprives students of valuable critical involvement in the writing process but also intensifies the challenges of measuring skills acquisition and enabling "complex plagiarism" (Perkins et al., 2019).

At the same time, outside of typical assessment tasks such as exams, essays, or written assignments, other tasks that are given as either formative or summative assessment in EAP may be vulnerable to overreliance on or exploitation of GenAI, such as writing annotated bibliographies, or even producing posts in a discussion forum or wiki. Even in low-stakes classroom activities, the use of AI translation or transcription software may lead to learners relying on such technologies to understand the meaning of texts, audiovisual material, peers, or instructors, meaning that less focus is given to engaging in unassisted skill development and communication in the target language.

Moreover, in cases where students have been able to use GenAI counter to the guidance of an instructor or institutional guideline, it may be impossible to prove that such usage has taken place. Although AI text detection software exists, research indicates that these tools are often inaccurate, easily circumvented (Chaka, 2023, 2024; Perkins et al., 2023; Perkins, Roe, et al., 2024; Weber-Wulff et al., 2023), and potentially biased against speakers of English as an Additional Language (EAL) (Liang et al., 2023). Furthermore, the line between a simple AI-powered writing assistant and an illegitimate overuse that obfuscates authorship is not always clear (Roe et al., 2023). This makes it difficult for educators to reliably identify AI-generated or AI-assisted work, further complicating the assessment process. Although there may be cases which are obvious, for example, a low-proficiency English learner submitting a piece of extremely sophisticated, well-written, and grammatically perfect research, with tell-tale 'tortured phrases' characteristic of GenAI writing (Cabanac et al., 2021), this in our estimation is likely to represent a relatively uncommon occurrence. Emerging GenAI technologies, such as deepfakes, pose additional ethical challenges for educational institutions, potentially impacting academic integrity, institutional trust, and student privacy (Roe & Perkins, 2024). These concerns highlight the need for a comprehensive ethical framework to guide the integration of AI and GenAI in educational settings.

Another element that draws attention to the need for shared, coherent frameworks for AI use is an evolving discourse in GenAI usage's acceptability, which has shifted from attempts to outright ban the use of such tools in education to a more nuanced acceptance of their value and potential benefits to teaching and learning. In some cases, bans in schools in the U.S. have been hastily repealed as more about GenAI's capabilities and use cases have come to light (Singer, 2023). In higher education, as GenAI tools have become increasingly ubiquitous and sophisticated, many universities and academic publishers are adapting their policies to permit their use, provided it aligns with principles of transparency and accountability (Perkins & Roe, 2023, 2024). The European Commission and United Nations Educational, Scientific, and Cultural Organization (UNESCO) have also provided guidance documents and policies which emphasise the potential benefits of GenAI in research and education (European Commission, 2024; Miao & Holmes, 2023). This shift acknowledges the fact that GenAI is unlikely to be a passing fad and instead represents a new and important technology that will be increasingly deployed in society at large.

Given these conflicting pressures and considerations, and a rapidly developing discourse, there is a need for ethically oriented, easily communicable frameworks for integrating AI tools into EAP, for both teaching and learning and assessment purposes. Such a framework must strike a balance between harnessing the potential benefits of these new technologies and teaching the critical evaluation of their uses and limitations. At the same time, it is of vital importance that assessment strategies accurately reflect learners' true abilities and skill development to maintain the integrity and quality of EAP instruction and outcomes.

With this in mind, this paper proposes an adaptation of the AI Assessment Scale (AIAS) (Furze et al., 2024; Perkins, Furze, et al., 2024) tailored specifically for EAP contexts. Our adapted scale, the EAP-AIAS, aims to provide a flexible tool that acknowledges both the opportunities and challenges presented by AI in EAP as well as a starting point for further study, investigation, and discussion on the role GenAI can play in pedagogy and assessment strategy. By offering a structured framework for AI integration, we hope to empower EAP practitioners to deal with this complex set of external pressures while maintaining the integrity and core objectives of EAP instruction. Within this framework, we refer to the scale as being most suited for assessment tasks, but at the same time, the flexibility of the framework would be suitable for all forms of EAP activity, including unassessed and formative assessment tasks given in class or for homework.

The remainder of this paper is structured as follows: First, we engage with the current literature on EAP, assessment, technology, and AI to provide a comprehensive background. We then present the adapted AI Assessment Scale (EAP-AIAS), explaining its levels and their applicability to common forms of assessment in EAP. Following this, we discuss the implementation of the EAP-AIAS through a series of potential use cases. Finally, we conclude with a discussion of the implications of our findings, acknowledgment of limitations, and proposals for future research directions.

## The Evolving Landscape of EAP and the Development of GenAI

English for Academic Purposes (EAP) has undergone significant evolution since its emergence as a distinct field in the 1960s. Originating in English for Specific Purposes (ESP), EAP has grown to represent a complex set of social practices within both institutional and global contexts (Charles, 2012). This evolution reflects the changing nature of higher education, the internationalisation of academia, and the growing recognition of the specific language needs of students engaging in English-medium instruction.

Today, the complexity and diversity of EAP contexts encompass a broad spectrum of learners and settings (Hyland & Jiang, 2021). EAP now caters to English as an Additional Language (EAL) students, undergraduate and postgraduate students, and first-language English speakers seeking to develop their academic language skills. It is practiced in various settings, including pre-sessional courses, embedded provision within academic programs, and standalone academic English programs.

Unlike general language acquisition courses, EAP's focus extends beyond solely developing linguistic macro-skills (reading, listening, writing, and speaking). Contemporary EAP focuses on the acquisition of the target language for use in specific contexts, for example, in lectures, presentations, and discussions, as well as (although to a lesser extent) study skills, such as techniques and strategies for note-taking or using Virtual Learning Environment discussion boards, and may involve English for General Academic Purposes (EGAP) or English for Specific Academic Purposes (ESAP) (Gillett, 2022). It may also involve teaching elements of culture (Gillett. 2022), and navigating culture-specific differences in learning approaches (Roe & Perkins, 2020). Additionally, EAP may focus on academic literacies, which involve understanding and producing texts within specific academic discourse communities (Lea & Street, 2006), and digital literacies, which are skills for navigating and utilising digital resources and technologies in academic contexts (Hafner, 2019).

The development of GenAI marks a pivotal moment in the teaching, learning, and research of English as a Foreign Language (EFL) (Pack & Maloney, 2023) and, by extension, EAP. Areas of practice that require attention include curriculum design, integrating AI literacy alongside traditional language and academic skills; assessment methods, developing new approaches to evaluate genuine student work in an AI-assisted environment; pedagogical approaches, adapting teaching methods to leverage AI tools while maintaining a focus on core EAP objectives; and ethical considerations, addressing issues of academic integrity, equity, and the appropriate use of AI in academic contexts.

As EAP is primarily a practical field that focuses on addressing real-life classroom challenges faced by teachers and students (Hyland & Jiang, 2021), GenAI represents a challenge that requires actionable, flexible, and practical solutions. There is increasing recognition of the need to enable students to engage in AI-mediated academic communication (Ou et al., 2024) while simultaneously ensuring that the fundamental goals of language acquisition and academic skill development are not compromised. At the same time, while there are many theoretical affordances for learners using GenAI in EAP, there are many potential limitations.

**Affordances and Limitations of GenAI in EAP**

The integration of Generative AI (GenAI) into English for Academic Purposes (EAP) contexts offers a range of potential benefits but also presents significant challenges and limitations. Understanding these affordances and constraints is crucial for developing an effective framework for AI use in EAP and identifying the potential affordances for students and educators.

GenAI tools, particularly advanced chatbots such as ChatGPT, Gemini, and Claude, can provide learners with opportunities for meaningful practice of the target language, potentially enhancing their overall proficiency (Sha, 2009). Research has indicated that interactions with AI chatbots can positively affect language proficiency by drawing attention to form and improving accuracy (Bibauw et al., 2019). Through sophisticated prompting, GenAI chatbots can play various roles in the learning process, acting as teaching, peer, or teachable agents, offering a flexible learning experience (Kuhail et al., 2023). Research has explored the use of prompt engineering with EFL learners to craft human-AI narratives, suggesting a potential innovation in teaching pedagogy (Woo et al., 2024).

Continuing with AI-powered chatbots, these GenAI applications can provide rich, varied inputs essential for second language acquisition (SLA). The ability to generate diverse examples and explanations can stimulate interest and contribute to overall language growth (Kohnke et al., 2023), and GenAI tools can generate personalised practice materials and facilitate real-time explanations, supporting diverse learning needs and allowing for the creation of tailored exercises that align with individual learner proficiency levels and learning styles.

In the context of academic writing, GenAI can assist with idea generation, vocabulary enhancement, structural guidance, feedback, and revision. These features can be particularly beneficial for developing writing skills by offering personalised support throughout the writing process (Xiao & Zhi, 2023). GenAI tools have also shown potential in promoting grit and resilience among L2 students, potentially helping them overcome learning challenges and thrive in their studies (Ghafouri, 2024).

However, the integration of GenAI into EAP presents considerable challenges. There is a risk that students will accept erroneous information produced by GenAI, which is particularly problematic in academic contexts where accuracy and reliability are paramount (Pack & Maloney, 2024). Indeed, GenAI tools mislead humans into believing they are intelligent, despite producing inaccurate, made-up, or unreliable content (Zhgenti & Holmes, 2023), and at the same time, the information produced by many AI tools is by nature culturally biased and representative of a Eurocentric worldview (Roe, 2024)

AI-generated text or audio content may also be above the proficiency level of many learners, potentially leading to overreliance on AI tools instead of developing their own language skills, which could lead to knowledge loss and a 'monoculture' of academic knowledge (Messeri & Crockett, 2024). As a result of the quality of outputs from GenAI tools, some have called for the teaching of evaluative judgment regarding AI output (Bearman et al., 2024). This also extends to the idea of using GenAI outputs for feedback, and the results of empirical studies suggest that automated writing evaluation from GenAI tools may have a place in English as a New Language (ENL) instruction (Escalante et al., 2023), and that GPT tools can be used for automated essay scoring with L2 writers (Mizumoto & Eguchi, 2023). Dai et al. (2023) was found to provide feedback that was more detailed than that of an instructor, yet showed high levels of agreement with the marking of the instructor. On the other hand, these systems are at an early stage in language assessment (Chiu et al., 2023), and questions remain over automated markings' ethics, legality, cost, and privacy (Kumar, 2023).

Equity and access issues arise, as not all students have equal access to the necessary technology or premium AI tools, potentially exacerbating existing educational disparities (Kohnke, 2023). Exploitation of workers and monopolistic domination of the GenAI market by companies in the Global North also raises broader questions about the ethics of GenAI in education and society (Zhgenti & Holmes, 2023).

Another question related to embedding GenAI in education is the additional work required by teachers, who must manage and integrate these tools into their teaching practices and stay updated with rapidly evolving AI technologies. In this space, not only does the integration of GenAI into EAP contexts necessitate a careful reconsideration of assessment practices, but also reconsideration of classroom approaches to equip students with the required skills set. While many of the affordances of GenAI in EAP relate to classroom activities and unassessed exploration, the role these tools should play in formal assessment is a subject of much discussion.

**Addressing GenAI in EAP Assessment**

Assessment has become an increasingly important topic in EAP, with bibliometric analysis showing an increasing trend in EAP assessment topics in scholarly publications from 1975 to 2019 (Charles, 2022). This growing focus reflects the recognition that assessment is a critical component of EAP pedagogy and program effectiveness. EAP assessment is inherently complex because of the intricate nature of language and academic skill evaluation (Bruce, 2015). This complexity arises from several factors, including the multidimensional nature of language proficiency, integration of language and content, genre-specific requirements, and cultural and linguistic diversity of EAP students. EAP assessments must evaluate not only linguistic competence but also the ability to use language appropriately in academic contexts, while also requiring students to demonstrate knowledge of academic subjects or research methodologies.

Several trends have emerged in EAP assessment in recent years. There is a growing shift from an exam culture to a learning culture, with an increasing emphasis on learner-oriented assessment and more formative and continuous assessment methods (Fazel & Ali, 2022). Authentic assessment has gained popularity, with a focus on tasks that reflect real-world academic activities (Uludag & McDonough, 2022). Portfolio-based assessments have also seen growing adoption, including a range of tasks such as summarising texts, creating research proposals, and giving oral presentations (Storch & Tapper, 2009). In addition, technology-enhanced assessment has become more prevalent, incorporating digital tools and platforms to allow for more diverse and interactive assessment tasks.

In this context, GenAI can help create more authentic assessment scenarios, but also raises concerns about academic integrity and the ability to accurately assess individual student capabilities. Traditional assessment methods need to be revised to effectively evaluate language and academic skills using a range of formats and modes of assessment, including those that allow for the critical use and evaluation of AI-generated content. Similarly, the potential for AI tools to provide instant, personalised feedback on assessments could assist the feedback process but raises questions about the quality and appropriateness of AI-generated commentary on student work.

Given these diverse challenges, any framework designed with the integration of GenAI into EAP must make assessment one of the main focal points. We argue that a specific EAP-AIAS framework is necessary to maintain assessment validity, communicate the appropriate use of GenAI to students and teachers, adapt to new technologies, and provide a balanced assessment profile. Furthermore, the implementation of the proposed EAP-AIAS should ensure that assessments remain valid measures of students' language and academic skills in an AI-enhanced environment, support the development of critical thinking and academic skills, and address issues of equity to prevent disadvantaging certain groups of students or exacerbating existing educational inequalities. We propose that the AIAS can be adapted for EAP to meet these goals and provide a suitable, flexible framework.

## The AI Assessment Scale (AIAS) and Its Adaptation for EAP

**Overview of the Original AIAS**

The AI Assessment Scale (AIAS) was originally conceived by Perkins, Furze, et al. (2024) to help educators in a wide range of disciplines adapt their assessments to the GenAI era. The original scale consists of five levels, ranging from 'No AI' to 'Full AI', designed to guide students on the appropriate use of GenAI tools while adhering to academic integrity principles. Results from a pilot study of the AIAS suggest that the use of the scale at an institutional level led to reduced cases of academic misconduct, improved student outcomes, and improved ethical integration of GenAI tools into assessment strategies (Furze et al., 2024). Various adaptations or translations of this scale have also been developed globally (Furze, 2024; Kılınç, 2024), demonstrating the potential value of similar tools.

While the original AIAS provides a valuable starting point, the unique needs of EAP necessitate a more tailored approach. EAP's dual focus on language proficiency and academic acculturation (Hyland, 2018) requires a framework that prioritises language and skill acquisition and ensures that the fundamental goal of developing genuine language skills is not compromised by AI use. Furthermore, an adapted AIAS framework needs to support academic skill development, foster criticality, promote AI literacy, and provide guidelines on the ethical use of AI tools.

**The EAP-Specific AI Assessment Scale (EAP-AIAS)**

Our proposed EAP-AIAS builds on the original five-point scale, redefining each level to align it with typical EAP tasks. By explicitly mapping each level to these established EAP approaches, the EAP-AIAS can provide a more robust and theoretically grounded framework for AI integration in EAP instruction.

| Level | Description | Focus | Example Tasks |
| --- | --- | --- | --- |
| **Level 1:** No AI Use | All language and skills tasks completed without AI assistance. | Developing core language skills and academic competencies independently. | Traditional examinations, in-class presentations, in-class comprehension and critical thinking tasks. |
| **Level 2:** AI-Assisted Language Input | AI used to generate or augment input materials. | Enhancing comprehension and analysis skills. | Inviting learners to engage with or create AI-generated texts for reading or listening comprehension and micro skills development. Instructor-created AI materials for assessment or practical use. |
| **Level 3:** AI for Limited Language or skills Practice | AI used for targeted practice of specific language and discourse features or academic skills development. | Reinforcing particular aspects of language, discourse, academic or discipline-specific conventions. | AI-generated content for controlled or semi-controlled practice of discipline-specific vocabulary and/or discourse features; simulated academic discussions with AI. |
| **Level 4:** AI-Assisted Task Completion with Critical Evaluation | Students use AI to assist in complex academic tasks but must critically evaluate and substantially revise AI outputs. | Developing critical thinking and digital literacy alongside language and academic skills. | Using AI for initial research or drafting, followed by substantial human revision, reflection, evaluation and critique. |
| **Level 5:** Selective AI Integration for Advanced Skills | More extensive AI use allowed but emphasizing its role in enhancing, not replacing, student work. | Preparing students for real-world academic and discipline-specific scenarios involving AI. | Using AI for data analysis in research projects, developing AI-enhanced academic presentations and discipline-specific outcome tasks |

*Table 1: The EAP-AIAS*

We recognise that the broad range of tasks present in EAP assessments means that these identified scale levels might be used at different times and for different purposes throughout EAP education, and that no level of the scale is inherently 'better' or more suitable than another. We have therefore adjusted the original colour scheme presented in the AIAS from a Red-Green scale to a more neutral palette of colours. This is to avoid any potential implicit suggestions to educators working with this scale that tasks or assessments with no AI usage are incorrect and that Full AI usage is the most appropriate or target level.

To illustrate the practical application of the EAP-AIAS, we present a series of possible use cases across different EAP contexts. These examples demonstrate how varying levels of the scale can be implemented at different levels of study and for different language learning objectives.

**Level 1: No AI Use**

In an intensive pre-sessional EAP program for international students, a Level 1 "No AI Use" approach could be applied for the final examination. This traditional assessment may be conducted in a secure, invigilated physical environment encompassing listening, reading, speaking, and writing components. By requiring students to demonstrate their EAP skills without the aid of AI or assistive technologies, this approach could provide a clear indication of their ability to independently apply language skills in their future academic studies. Such examinations might offer a baseline measure of language competency unaffected by AI or other technological interventions, aligning with the core principles of Level 1 in the EAP-AIAS.

**Level 2: AI-Assisted Language Input**

For first-year undergraduate students in an in-sessional English for Specific Academic Purposes (ESAP) program focusing on business subjects, a Level 2 "AI-Assisted Language Input" approach could be implemented. GenAI chatbots may be utilised to produce discipline-specific graded readings, aligning with the EAP-AIAS guidelines for AI-assisted idea generation and structuring. Instructors could guide students to generate texts at appropriate difficulty levels, potentially encouraging them to experiment with the chatbot to explain key terms or interpret meanings outside class time. This Level 2 approach may stimulate learner interest by offering opportunities to acquire new discipline-specific vocabulary in context and familiarise students with genre conventions while ensuring that the final work remains AI-free.

**Level 3: AI for Limited Language Practice**

In a postgraduate EAP module focusing on academic writing, a Level 3 "AI for Limited Language Practice" approach could be adopted. Students might be encouraged to use GenAI applications such as ChatGPT, Claude, or Gemini to develop their sentence and paragraph structure skills in line with the EAP-AIAS guidelines for AI-assisted editing. The process could involve creating discipline-specific word lists, collocations, or clichés based on genre requirements, which could then be incorporated into writing tasks. Students could input their written work into the GenAI application for initial feedback on lexical, grammatical, and structural elements. They might then be tasked with analysing and justifying any changes made based on AI suggestions, potentially promoting critical engagement with the tool. This Level 3 approach could improve students' use of discipline-specific languages, with the critical analysis of AI grammar suggestions possibly leading to a deeper understanding of linguistic structures.

**Level 4: AI-Assisted Task Completion with Critical Evaluation**

In a postgraduate EAP course on research writing, a Level 4 "AI-Assisted Task Completion with Critical Evaluation" approach could be implemented to develop research proposals. This would align with the EAP-AIAS framework for AI task completion with human evaluation. The process may begin with an introduction to the UNESCO (Miao & Holmes, 2023) and COPE (2023) guidelines on the ethical use of GenAI in research. Students could then use AI tools to generate initial ideas and create outlines for their research proposals. A critical component of this Level 4 task might be the evaluation of AI-generated content, which students would need to revise substantially and justify their changes. Peer review sessions could focus on discussing how effectively students have adapted and improved upon AI-generated content. Final submissions may include a reflection on the process of working with and critiquing AI-generated material, fulfilling the critical evaluation aspect of Level 4.

**Level 5: Selective AI Integration for Advanced Skills**

In an EAP workshop for doctoral students, a Level 5 "Selective AI Integration for Advanced Skills" approach could be employed, representing the highest level of AI integration in the EAP-AIAS framework. GenAI tools can be used extensively to assist in synthesising large volumes of research for literature review. Students could be guided to use these tools in accordance with COPE best practices (2023) and UNESCO guidelines (Miao & Holmes, 2023). The process might involve prompting AI to summarise key points from multiple papers, followed by a critical evaluation of these summaries. In line with Level 5 of the EAP-AIAS, students could be required to identify gaps in AI-generated syntheses and develop their own arguments based on a more comprehensive understanding of the field. Final presentations may include a critical reflection on the role of AI in the research process and its limitations, potentially demonstrating the advanced skills and critical thinking fostered by the highest level of AI integration.

## Discussion

The advantage of the AIAS in EAP is that it provides a structured yet flexible approach to GenAI integration, providing clear guidelines for usage in specific tasks. The EAP-AIAS can help empower educators to effectively incorporate AI tools into their teaching and assessment practices while ensuring that students develop the necessary language skills which can be evaluated through a range of tasks. The EAP-AIAS may be used to help maintain academic integrity by educating students about acceptable use where it is appropriate and providing a system for communication. If students are given a clear framework which clarifies the acceptable use cases of GenAI and receive encouragement on adhering to values of transparency and ethics, then the EAP-AIAS can help to guide the successful integration of GenAI.

However, successful implementation of the EAP-AIAS requires careful consideration of several factors. First, instructors should become comfortable with each level before moving to the next. Clear guidelines with explicit instructions and examples should be provided for each level to clarify what constitutes appropriate GenAI use. As part of their effective use, instructors need to have comprehensive training on both the technical aspects of AI tools and their pedagogical applications.

The EAP-AIAS represents a potential framework that may benefit EAP pedagogy by explicitly addressing AI use while reaffirming the importance of foundational language and academic skills. This scale aligns with calls for enhanced digital literacy in EAP, providing a structured approach to develop these skills alongside traditional language competencies. EAP programmes with an explicit focus on digital literacy contribute to better academic integrity and assist with better accessing the course content (Roche, 2017).

The emphasis on critical evaluation of AI-generated content, particularly in types of tasks relating to levels 4 and 5, resonates with critical EAP approaches, encouraging students to develop evaluative judgement with both content and tools, thus fostering criticality.

It also contributes to the development of metacognitive skills, going beyond high-order thinking skills and allowing space for developing reflective thinking. In this way, the EAP-AIAS serves as a bridge between traditional EAP instruction and the demands of an increasingly AI-influenced academic environment.

Questions remain regarding the long-term impact of AI integration on language and skill acquisition. The principle of maintaining a majority of instruction and classroom interaction at Level 1 (No AI Use) appears crucial, but longitudinal studies are needed to confirm its effectiveness. With the influence of AI, academic skills development is likely to go through significant changes - certain subskills might become obsolete, which will require adaptation of core methodologies and full GenAI integration in the EAP curriculum with design of a relevant set of assessment tools, possibly AI –powered assessment tools such as EAP Talk – an AI software for assessing academic speaking skills (Wang & Zou, 2023).

The issue of academic integrity in AI-enhanced environments remains complex. Our experience suggests that fostering a culture of ethical GenAI use through explicit instruction and reflection is essential. The potential of AI tools to exacerbate existing educational inequalities must be carefully monitored and addressed. The aim is to create a constructive learning environment in an EAP classroom where students benefit from using AI tools in developing their academic skills while learning to use them ethically and maintaining academic integrity. The scale may need further refinement to address discipline-specific needs within EAP, as the appropriate level of AI integration may vary across academic fields and discourses. The current scale may not fully capture the nuances of AI use in collaborative tasks, an important aspect of academic work that requires further exploration. The fast-paced evolution of GenAI technologies indicates that the scale requires regular updates to remain relevant.

Several avenues for future research emerge from our implementation of the EAP-AIAS. These include conducting longitudinal studies tracking students' language and skills development and academic performance as they progress through the EAP-AIAS levels over extended periods; comparative studies examining learning outcomes between EAP-AIAS-guided instruction and traditional EAP approaches; investigating the role of AI in developing specific EAP skills, such as academic listening and speaking;

exploring how the EAP-AIAS might be adapted for discipline-specific EAP courses; and examining the most effective methods for developing critical AI literacy within EAP contexts.

Based on our findings, we offer several recommendations to EAP practitioners and institutions. These include beginning with a phased implementation, focusing on one or two levels of the scale initially, investing in comprehensive professional development programs for EAP instructors, developing clear institutional policies on AI use in academic work aligned with the principles of the EAP-AIAS, establishing support structures for students, including workshops on ethical AI use and access to necessary tools, and regularly reviewing and adapting the implementation of the EAP-AIAS based on feedback from instructors and students.

## Conclusion

The EAP-AIAS represents a significant step towards addressing the challenges and opportunities presented by GenAI in EAP instruction. Although further research and refinement are necessary, the scale provides a valuable framework for integrating GenAI tools in a manner that enhances, rather than undermines, the core objectives of EAP. As we move forward, maintaining a balance between embracing technological advancements and preserving the fundamental principles of language acquisition and academic skill development is crucial for shaping the future of EAP pedagogy.

This scale offers a structured yet flexible approach to EAP assessment in the GenAI era, empowering both educators and learners to engage critically and ethically with GenAI technologies in academic contexts. The journey of integrating GenAI into EAP is ongoing, and the EAP-AIAS should be viewed as a living framework, open to adaptation and refinement as our understanding of GenAI's role in language education evolves. By continuing to engage critically with these technologies, conduct rigorous research, and prioritise the core goals of EAP, we can ensure that GenAI becomes a valuable tool in our pedagogical arsenal rather than a disruptive force that undermines the essence of language and academic skills development.

**AI Usage Disclaimer**

This study used Generative AI tools (Claude 3.5 Sonnet) for revision and editorial purposes throughout the production of the manuscript. The authors reviewed, edited, and take responsibility for all outputs of the tools used in this study